\documentclass[aps,pra,preprint,superscriptaddress,12pt]{revtex4-2}

\usepackage{amsmath}
\usepackage{hyperref}
\usepackage{graphicx}
\usepackage{xcolor}
\usepackage{epstopdf}
\usepackage{verbatim}
\usepackage{physics}
\usepackage[normalem]{ulem}
\usepackage{siunitx}
\usepackage{orcidlink}
\begin{document}
\title{Probing electronic state-dependent conformational changes in a trapped Rydberg ion Wigner crystal}

\author{Marion Mallweger~\orcidlink{0009-0009-8142-2988}}
\email[]{marion.mallweger@fysik.su.se}
\affiliation{Department of Physics, Stockholm University, 10691 Stockholm, Sweden}
\author{Natalia Kuk~\orcidlink{0009-0003-0601-0314}}
\affiliation{Department of Physics, Stockholm University, 10691 Stockholm, Sweden}
\author{Vinay Shankar~\orcidlink{0009-0005-6551-7689}}
\affiliation{Department of Physics, Stockholm University, 10691 Stockholm, Sweden}
\author{Robin Thomm~\orcidlink{0009-0000-8105-8690}}
\affiliation{Department of Physics, Stockholm University, 10691 Stockholm, Sweden}
\author{Harry Parke~\orcidlink{0000-0001-6120-5470}}
\affiliation{Department of Physics, Stockholm University, 10691 Stockholm, Sweden}
\author{Ivo Straka~\orcidlink{0000-0003-2675-6335}}
\affiliation{Department of Physics, Stockholm University, 10691 Stockholm, Sweden}
\author{Weibin Li~\orcidlink{0000-0001-6731-1311}}
\affiliation{School of Physics and Astronomy and Centre for the Mathematics
and Theoretical Physics of Quantum Non-Equilibrium Systems, University of Nottingham, Nottingham, NG7 2RD, United Kingdom}
\author{Igor Lesanovsky~\orcidlink{0000-0001-9660-9467}}
\affiliation{School of Physics and Astronomy and Centre for the Mathematics
and Theoretical Physics of Quantum Non-Equilibrium Systems, University of Nottingham, Nottingham, NG7 2RD, United Kingdom}
\affiliation{Institut f\"ur Theoretische Physik and Center for Integrated Quantum Science and Technology, Universit\"at T\"ubingen, Auf der Morgenstelle 14, 72076 T\"ubingen, Germany}
\author{Markus Hennrich~\orcidlink{0000-0003-2955-7980}}
\email[]{markus.hennrich@fysik.su.se}
\affiliation{Department of Physics, Stockholm University, 10691 Stockholm, Sweden}

\begin{abstract}
State-dependent conformational changes play a central role in molecular dynamics, yet they are often difficult to observe or simulate due to their complexity and ultrafast nature. One alternative approach is to emulate such phenomena using quantum simulations with cold, trapped ions. In their electronic ground state, these ions form long-lived Wigner crystals. When excited to high-lying electronic Rydberg states, the ions experience a modified trapping potential, resulting in a strong coupling between their electronic and vibrational degrees of freedom. In an ion crystal, this vibronic coupling creates electronic state-dependent potential energy surfaces that can support distinct crystal structures -- closely resembling the conformational changes of molecules driven by electronic excitations. Here, we present the first experimental observation of this effect, by laser-coupling a single ion at the centre of a three-ion crystal to a Rydberg state. By tuning the system close to a structural phase transition, the excitation induces a state-dependent conformational change, transforming the Wigner crystal from a linear to a zigzag configuration. This structural change leads to a strong hybridisation between vibrational and electronic states, producing a clear spectroscopic signature in the Rydberg excitation. Our findings mark the first experimental step towards using Rydberg ions to create and study artificial molecular systems.
\end{abstract}

\maketitle
\newpage
\section{Introduction}
Electronic transitions in molecules are fundamental to chemical reactions and biological processes \cite{dimitriev_dynamics_2022, ponseca_ultrafast_2017, brunk_mixed_2015,gozem_theory_2017}. A prime example of this is vision, in which a photon-induced electronic excitation of a molecule causes a conformational change, leading to stimulation of the optical nerve \cite{rinaldo2014, Schoenlein1991}. These transitions occur on timescales of femto- to picoseconds and Ångström length scales, making direct observation of the dynamics challenging. Theoretical modelling is also limited, despite the fact that the molecular constituents and their interactions are well understood. This is due to the exceedingly high-dimensional Hilbert space spanned by the electronic and nuclear degrees of freedom, which renders ab initio numerical simulations extremely demanding. While well-established adiabatic approximations are available \cite{koppel_theory_1981,gonzalez_progress_2012}, they typically break down near intersecting potential energy surfaces (PESs), with the aforementioned process of vision being one example. These challenges create a need for a quantum simulator that can mimic the dynamics of coupled electrons and nuclei on directly accessible time and length scales. A promising platform in this regard are Wigner crystals formed by trapped ions. They combine internal electronic states with external collective vibrational modes. Recently, such a system was employed for the first direct measurement of a geometric phase near conically intersecting PESs \cite{whitlow_quantum_2023,valahu_direct_2023}. Moreover, structural transitions in a Wigner crystal have been reported as a result of the photoionisation of an ion to a doubly charged state \cite{feldker_mode_2014}. This is reminiscent of an electronic state-dependent conformational change, albeit irreversible, of a large molecule.\\

Building on these ideas, this work considers the laser excitation of a single ion embedded in a three-ion Wigner crystal to an electronically high-lying Rydberg state \cite{mokhberi_chapter_2020, Higgins2017Jun, Feldker2015Oct}. Due to the high polarisability of this electronic state, the excited ion experiences a local modification of its trapping potential \cite{Higgins2019Oct}. Near a conformational change of the Wigner crystal, this results in a strongly enhanced coupling between the vibrational and electronic degrees of freedom \cite{Weibin2011,gambetta_exploring_2021,gambetta_long-range_2020}.
This so-called vibronic coupling renders the structure of the Wigner crystal dependent on the electronic state. We present experimental evidence of this phenomenon by studying a three-ion crystal of $^{88}$Sr$^+$, which changes from a linear to a zigzag configuration when the central ion is excited to the Rydberg state. This transition leaves a clear signature in the Rydberg excitation spectrum, which further allows us to demonstrate that the structure of the Rydberg excited Wigner crystal can be shaped with the help of a microwave field. The tunability of such a setup offers the potential to directly engineer PES for the quantum simulation of molecular processes.

\section{Electronic state-dependent Wigner crystals}
Trapped ions self-assemble to form a Wigner crystal if they are confined in the same potential and are cooled to sufficiently low temperatures \cite{Walther1995Jan}. The conformation of this crystal is controlled by the interplay between confinement and Coulomb repulsion. Linear Paul traps, which use static and oscillating electric quadrupoles to confine ions, are typically operated in a regime in which linear crystals form. However, changing the ratio of the radial to the axial confinement strength, parametrised by the trap frequencies, can result in a conformational change from a linear to a zigzag (ZZ) shaped Wigner crystal \cite{rainzen1992}. For a crystal of $N$ ions and a fixed axial trap frequency $\omega_\mathrm{z}$, this change occurs at the critical radial trap frequency \cite{schiffer1993}
\begin{align}\label{eq:anisotropy}
    \omega_\mathrm{x,c}=
    0.81\,\omega_\mathrm{z}\,N^{0.87}.
\end{align}
In addition to their charge, trapped ions interact with the confining electric field through their induced dipole moment. While this interaction plays only a minor role in their electronic ground state, it becomes significant when high-lying Rydberg states with a large principal quantum number $n$ are excited. Such states experience strong spatially dependent energy shifts, which are dependent on the polarisability $\mathcal{P}_\mathrm{r}\propto n^7$ \cite{Higgins2019Oct, Niederlander2023Mar}. The state-dependent coupling to the electric fields of the Paul trap directly translates to a state-dependent confinement of the ion, which can have a major impact on the conformation of the Wigner crystal. As theoretically predicted in~\cite{Weibin2011}, the Rydberg excitation of even a single ion can induce a conformational change between a linear and a ZZ crystal. Specifically, in a three-ion crystal, of which the central ion is excited to the Rydberg state, the critical radial frequency (with respect to the ground state trapping frequency) is given by \cite{Weibin2011}
\begin{align}\label{eq:wchange}
\omega^{(\mathrm{r})}_\mathrm{x,c}\approx\sqrt{\omega_\mathrm{x,c}^2 + \frac{2e^2\alpha^2+4e^2\beta^2}{M}\mathcal{P}_\mathrm{r}}.
\end{align}
Here, $\alpha$ and $\beta$ denote the trapping field gradients of the oscillating and static quadrupole fields, respectively, $e$ is the elementary charge and $M$ is the ion mass. The sign of $\mathcal{P}_\mathrm{r}$ determines whether the Rydberg excitation of a single ion results in stronger ($\mathcal{P}_\mathrm{r}<0$) or weaker ($\mathcal{P}_\mathrm{r}>0$) critical radial frequencies of the ground-state ion crystal. For radial trapping frequencies of $\omega^{(\mathrm{r})}_\mathrm{x,c}>\omega_\mathrm{x}>\omega_\mathrm{x,c}$, the stable conformation of the Wigner crystal is linear, when all ions are in the ground state and ZZ when the central ion is excited to the Rydberg state.

This phenomenon is reminiscent of a conformational change of a molecule upon electronic excitation. For example, in the process of vision, the rhodopsin chromophore undergoes a conformational change from \emph{cis} to \emph{trans}, when promoted from its electronic ground state PES to a high-lying one under the absorption of a photon \cite{rinaldo2014, Schoenlein1991}.

\section{Spectral signatures of vibronic coupling}
To demonstrate the state-dependent conformational change of a Wigner crystal in our experiment, we use a linear chain of three $^{88}$Sr$^+$ ions confined in a linear macroscopic Paul trap. The central ion is excited from the ground state $\ket{g} \equiv \ket{4\,^2D_{5/2}, m_J=-\frac{5}{2}}$ to the highly excited state $\ket{r}\equiv\ket{46\,^2S_{1/2}, m_J=-\frac{1}{2}}$ with a principal quantum number $n=46$ and $\mathcal{P}_\mathrm{46S}>0$. We drive the transition $\ket{g}\leftrightarrow\ket
r$ via an intermediate level $\ket{P}\equiv\ket{6\,^2P_{3/2}, m_J=-\frac{3}{2}}$, see Supplemental Material (\textbf{SM}) and \cite{higgins_quadrupol_2021, mokhberi_chapter_2020,bao_microwave-dressing_2025}. The state-dependent polarisability creates a situation in which the PES of the ground state and the highly excited state differ significantly. Concretely, this means that the ion crystal conformation as well as the vibrational mode spectrum become dependent on the electronic state.
\begin{figure}[t!]
\centering
\includegraphics[width=\textwidth]{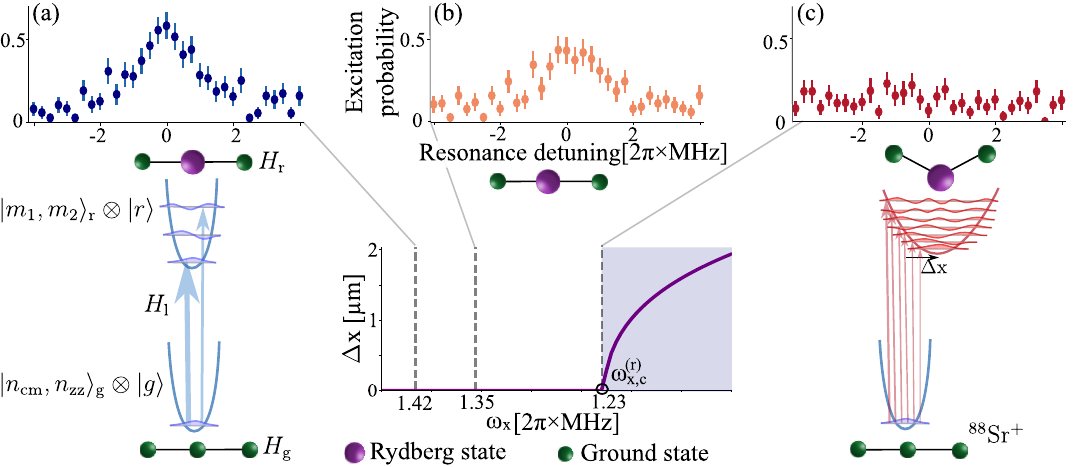}
\caption{\textbf{Spectral signature of conformational change.} The central ion of a linear three-ion crystal is laser excited to a Rydberg state. We measure the excitation probability as a function of the radial trap frequency $\omega_\mathrm{x}$ and the detuning from the resonance. (a) For $\omega_\text{x}\gg\omega_\mathrm{x,c}^\mathrm{(r)}$ , the PESs of the ground state and Rydberg state crystal are similar. The FC factors are large for the nearly identical phonon states (indicated by the thick arrow), but almost zero if they are different (thin arrows). The Rydberg excitation is largely independent of the phonon dynamics. The excitation probability (top panel) is high and a clear Rydberg resonance line is visible. (b) When $\omega_\text{x}$ approaches $\omega_\mathrm{x,c}^\mathrm{(r)}$, the PES is changed by the Rydberg polarisability $\mathcal{P}_{\text{r}}$, such that the motional modes of the ground and Rydberg state will differ more significantly. The ensuring vibronic coupling reduces the height of the spectral line. (c) When the conformational change from a linear ion chain to a zigzag configuration occurs, the potential minima of the two electronic state-dependent PESs are displaced by a distance $|\Delta \text{x}|$. Combined with a significant change in the trap frequencies, this leads to a broad distribution of FC factors. As a result, the Rydberg resonance line disappears. The radial trap frequencies are (a) $\omega_\mathrm{x}=2\pi\times\SI{1.42}{\mega\hertz}$, (b) $\omega_\mathrm{x}=2\pi\times\SI{1.35}{\mega\hertz}$ and (c) $\omega_\mathrm{x}=2\pi\times\SI{1.23}{\mega\hertz}$ in the three experimental datatsets. The critical radial trap frequency is $\omega_\mathrm{x,c}^\mathrm{(r)}=2\pi\times\SI{1.23}{\mega\hertz}$ and the axial frequency is $\omega_\mathrm{z}=2\pi\times\SI{0.778}{\mega\hertz}$. Error bars represent quantum projection noise (68\% confidence intervals).}
\label{fig:Fig_1}
\end{figure}

This has dramatic consequences on the laser coupling to the Rydberg state. The excitation process, illustrated in Fig.~\ref{fig:Fig_1}, is modelled through the Hamiltonian $H =  H_\mathrm{g} \otimes \lvert g\rangle \langle g\rvert + H_\mathrm{r} \otimes \lvert r\rangle \langle r\rvert+ H_\mathrm{l}$. The first term describes the vibrational dynamics of the crystal, which is associated with the electronic state $\ket{g}$. The corresponding PES is given by a combination of the confining electric field and the Coulomb interaction between the ions. For a three-ion crystal, this gives rise to three distinct collective phonon modes in each direction. However, only two of the radial modes are relevant in the Rydberg excitation process (see \textbf{SM} for details). These modes are the centre-of-mass (CM) and ZZ motional mode, whose dynamics are governed by the Hamiltonian $H_\mathrm{g} = \omega_\mathrm{cm} a_\mathrm{cm}^{\dagger}a_\mathrm{cm} +\omega_\mathrm{zz}a_\mathrm{zz}^{\dagger}a_\mathrm{zz} + (\omega_\mathrm{cm}+\omega_\mathrm{zz})/2$, ($\hbar \equiv 1$). Here, $\omega_j$ and $a_j^{\dagger}$ ($a_j$) are the mode frequency and creation (annihilation) operator of the $j$-th mode ($j=\mathrm{cm,zz}$). The Hilbert space of the two modes is spanned by the number state $\ket{\mathbf{n}}_\mathrm{g}=\ket{n_\mathrm{cm},n_\mathrm{zz}}_\mathrm{g}$, where $n_\mathrm{cm}$ and $n_\mathrm{zz}$ are the phonon numbers in the CM and ZZ mode. The second term of $H$ describes the vibrational motion when the central ion is excited to the Rydberg state. The corresponding PES is not only formed by the trapping potential and the Coulomb interaction, but also by the coupling of the induced dipole moment to the electric field of the Paul trap. The resulting energy shift, which is governed by $\mathcal{P}_\mathrm{r}$, leads to a modified shape of the PES, resulting in different motional eigenmodes in the Rydberg state. These two modes have frequencies $\omega_\mathrm{1}^{\mathrm{(r)}}$ and $\omega_\mathrm{2}^{\mathrm{(r)}}$, and differ from $\omega_\mathrm{cm}$ and $\omega_\mathrm{zz}$ (see \textbf{SM}). Their Hamiltonian reads $H_\mathrm{r}= \omega^{\mathrm{(r)}}_{1}b_\mathrm{1}^{\dagger}b_\mathrm{1} + \omega^{\mathrm{(r)}}_{2}b_\mathrm{2}^{\dagger}b_\mathrm{2} + (\omega^{\mathrm{(r)}}_{1}+\omega^{\mathrm{(r)}}_{2})/2$, with the corresponding creation (annihilation) operators $b_j^{\dagger}$ ($b_j$) of the two phonon modes of the Rydberg state PES. Their Hilbert space is spanned by the phonon number states $\ket{\mathbf{m}}_\mathrm{r}=\ket{m_\mathrm{1},m_\mathrm{2}}_\mathrm{r}$ with $m_\mathrm{1,2}$ denoting the phonon numbers. The last term of $H$ describes the laser coupling between the ground state and the Rydberg state of the central ion, which is described by the Hamiltonian 
\begin{equation}
    H_\mathrm{l} = \delta\,  \mathbf{I}_p\otimes \lvert r\rangle \langle r\rvert +\frac{\Omega}{2}\left(\sum_{\mathbf{n,m}}C_{\mathbf{n}}^\mathbf{m}\ket{\mathbf{n}}_\mathrm{g} \bra{\mathbf{m}}_\mathrm{r}\otimes \lvert g\rangle \langle r\rvert + \mathrm{h.c.}\right).
\end{equation}
Here, $\delta$ is the detuning between the atomic resonance frequency and the laser frequency, and $\Omega$ is the Rabi frequency that couples states $\ket{g}$ and $\ket{r}$. We have defined $\mathbf{I}_p$ to be the identity operator of the total phonon Hilbert space. 

The laser excitation from the ground to the Rydberg state depends on the Franck-Condon (FC) factors $|C_{\mathbf{n}}^{\mathbf{m}}|^2$, which are determined by the overlap of the motional wave functions, $C_{\mathbf{n}}^{\mathbf{m}}=C_{n_\mathrm{cm},n_\mathrm{zz}}^{m_\mathrm{1},m_\mathrm{2}}= {_\mathrm{g}\langle} n_\mathrm{cm},n_\mathrm{zz}|m_\mathrm{1},m_\mathrm{2}\rangle_\mathrm{r}$~\cite{BorrelliRaffaele2013Feb}. In the regime far away from the conformational change, the linear Wigner crystal of ground state ions will stay in a linear configuration upon Rydberg excitation. Hence, the motional wave functions within the two PESs are similar, and therefore the FC factors are large when $\mathbf{n}=\mathbf{m}$, and close to zero otherwise. This is illustrated in Fig.~\ref{fig:Fig_1}(a) where the transition strength between the electronic and vibrational states is indicated by the thickness of the arrow. As a result, vibrational state changes under laser excitation are suppressed and a clear spectral line is visible, centred at the expected resonance frequency, i.e., $\delta =0$. When $\omega_{\text{x}}$ is lowered towards the critical frequency $\omega_{\text{x,c}}^{(\text{r})}$, the PESs of the ground and the Rydberg states start to differ significantly. The vibrational eigenstates of the ground state PES have overlap with multiple vibrational eigenstates of the Rydberg state PES (see \textbf{SM} for illustration). There are many non-zero FC factors, and hence the transition strength gets distributed over many frequencies (as indicated by the arrows). This leads to a less pronounced spectral line, as seen in Fig.~\ref{fig:Fig_1}(b). Finally, when $\omega_{\text{x,c}}<\omega_{\text{x}}<\omega_{\text{x,c}}^{(\text{r})}$, the linear Wigner crystal will undergo a conformational change to a ZZ configuration as the central ion is excited to the Rydberg state. The coupling to the Rydberg state drastically modifies the landscapes of the PESs such that their potential minima are displaced, i.e.~$|\Delta x|\neq 0$, see Fig.~\ref{fig:Fig_1}(c). As a result, many FC factors are non-zero and small (see \textbf{SM}). In addition, the phonon state energies are mismatched due to the displacement $\Delta x$, such that the Rydberg excitation is shifted out of resonance. The combination of these effects becomes visible as a suppression of the Rydberg resonance shown in Fig.~\ref{fig:Fig_1}(c). Moreover, all three ions are displaced from the electric field null of the trapping potential in the ZZ crystal. The strong off-axis oscillating electric field experienced by the ions causes excess micromotion, further reducing the overall excitation probability for the Rydberg transition \cite{higgins_coherent_2017,martins_impact_2024}. 

\section{Control of potential energy surfaces}
\begin{figure}[t]
\centering
\includegraphics[width=\textwidth]{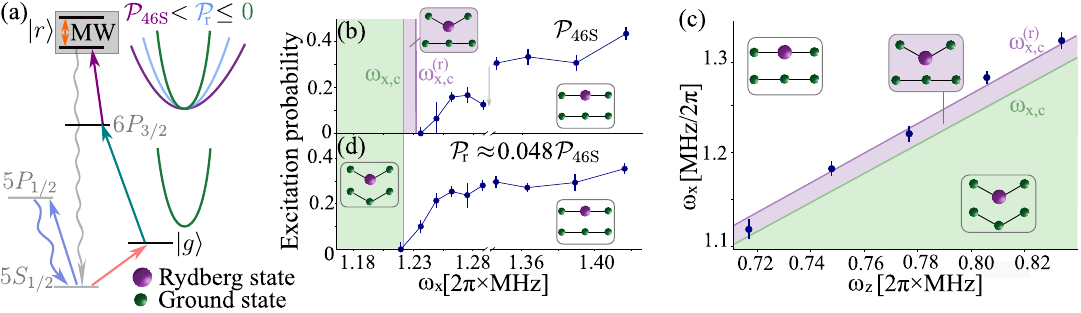}
\caption{\textbf{Potential energy surface control.} (a) Dressed Rydberg states $\ket{r}$, created via MW coupling, exhibit tunable polarisability. This enables precise control over the shape of the state-dependent PESs. (b) The central ion is coupled to the Rydberg state $\ket{46S}$ with a negative polarisability $\mathcal{P}_\mathrm{46S}$. Due to different PESs in ground and Rydberg states, the excitation probability decreases as the conformational change is approached, leading to a drop, indicated by the arrow, between the datasets for different radial trapping frequencies. (c) Phase diagram of the conformational change of a Wigner crystal induced by excitation of the Rydberg state $\ket{46S}$ for different trapping frequency ratios. (d) The central ion is excited to a dressed Rydberg state with reduced polarisability value. This suppresses the change in PES between the ground and excited state, leading to a recovery of the Rydberg resonance. The state-dependent conformational change of the Wigner crystal can be strongly suppressed. For all scans shown here, $\omega_\mathrm{z}=2\pi\times\SI{0.778}{\mega\hertz}$. Error bars represent quantum projection noise (68\% confidence intervals).}
\label{fig:Fig_2}
\end{figure}

Here, we demonstrate a method for controlling the PES by engineering Rydberg states with tunable polarisabilities. By varying the polarisability $\mathcal{P}_\mathrm{r}$, we can shape the PES in the Rydberg state so that it more closely matches that of the ground state, as shown in Fig.~\ref{fig:Fig_2}(a). This control is characterised spectroscopically by its impact on the excitation probability, which depends sensitively on the conformation of the Wigner crystal. In this way, we use the conformation-dependent excitation signal to probe the influence of $\mathcal{P}_\mathrm{r}$.

Before tuning $\mathcal{P}_\mathrm{r}$, we first characterise the conformational change of the Wigner crystal through the laser excitation of a single Rydberg state $\ket{46S}$ with a fixed polarisability $\mathcal{P}_\mathrm{46S}$. Fig.~\ref{fig:Fig_2}(b) shows the excitation probability at fixed axial confinement, as the radial trapping frequency is varied towards the critical radial trapping frequency $\omega_\mathrm{x,c}^\mathrm{(r)}$. As the radial trapping frequency decreases, the overlap between the state-dependent PESs is reduced, leading to a decrease in excitation probability. This drop becomes particularly pronounced close to the critical frequency, as indicated by the arrow in Fig.~\ref{fig:Fig_2}(b). Once the conformational change occurs at the critical frequency $\omega_\mathrm{x,c}^\mathrm{(r)}$, the Rydberg resonance in Fig.~\ref{fig:Fig_2}(b) disappears, reflecting the negligible overlap between the vibrational number states of the two PESs and the displacement of the PES of the Rydberg state. Repeating the scan for different axial trap frequencies allows us to map out the boundary of the conformational transition, shown as blue data points in Fig.~\ref{fig:Fig_2}(c). The disappearance of the resonance aligns well with the predicted critical value of the axial trapping frequency $\omega_\mathrm{x,c}^\mathrm{(r)}$, calculated using Eq.~(\ref{eq:wchange}).

Next, we tune the PES of the Rydberg state. The difference in PESs between ground and Rydberg states depends on the polarisability. By significantly reducing $\mathcal{P}_\mathrm{r}$, the PESs become almost state independent. We realise this by coupling the two Rydberg states $\ket{46S}$ and $\ket{46P}\equiv\ket{46{}^2P_{1/2}\textbf{ }, m_J=+\frac{1}{2}}$ using a microwave (MW) field, as shown in Fig.\ref{fig:Fig_2}(d). The resulting dressed state $\ket{r} = \cos(\theta) \ket{46S} - \sin(\theta) \ket{46P}$ is a superposition between the two Rydberg states with a mixing ratio given by $\theta$ \cite{pokorny2020}. The total polarisability of the dressed state reads $\mathcal{P}_\mathrm{r}=\mathcal{P}_\mathrm{46S}\sin^2{\theta}+\mathcal{P}_\mathrm{46P}\cos^2{\theta}$. Since $\mathcal{P}_\mathrm{46S}$ and $\mathcal{P}_\mathrm{46P}$ have opposing signs, $\mathcal{P}_\mathrm{r}$ can be tuned to almost zero \cite{pokorny2020} (for more details see \textbf{SM}). The effect of reducing the polarisability, and thereby minimising the difference between the ground and Rydberg state PESs, is depicted in Fig.~\ref{fig:Fig_2}(d). This leads to $\omega_\mathrm{x,c}^\mathrm{(r)}\approx\omega_\mathrm{x,c}$.

To demonstrate the control over the state-dependent PESs we use a dressed state with significantly reduced polarisability $\mathcal{P}_\mathrm{r}\approx0.048\mathcal{P}_\mathrm{46S}$. The excitation probability near the conformational change of the Wigner crystal is shown in Fig.~\ref{fig:Fig_2}(d). As in Fig.~\ref{fig:Fig_2}(b), the axial trap frequency was fixed while the radial trap frequency is varied. Due to the reduced polarisability, the dressed state shows a more consistent excitation probability over a wider range of trap frequencies, including a non-zero excitation probability at the critical trap frequency where the excitation previously vanished for Rydberg states $\ket{46S}$. Remarkably, near the conformational transition, the excitation probability for the dressed Rydberg state exceeds that of the Rydberg state $\ket{46S}$, despite the fact that the excitation probability of the dressed state is reduced by a factor $\cos^2(\theta)$ compared to the state $\ket{46S}$. The remaining decrease of the excitation probability takes place as the conformational change of the ground state ion crystal is approached. This indicates that the reduced polarisability causes the critical radial trap frequency for the dressed Rydberg state $\omega_\mathrm{x,c}^\mathrm{(r)}$ to approach that of the ground state $\omega_\mathrm{x,c}^\mathrm{(r)}\approx\omega_\mathrm{x,c}$. While Fig.~\ref{fig:Fig_2}(b) and Fig.~\ref{fig:Fig_2}(d) showcase two extreme examples, $\mathcal{P}_\mathrm{r}$ can be freely set between $\mathcal{P}_\mathrm{46S}$ and $\mathcal{P}_\mathrm{46P}$ by varying the detuning of the microwave field (see \textbf{SM} for more detail). This makes the critical frequency $\omega_\mathrm{x,c}^\mathrm{(r)}$ tunable.

\section{Summary and outlook}
We have conducted the first experimental investigation of a conformational change in a Wigner crystal via Rydberg excitation. The underlying mechanism relies on electronic state-dependent PESs that govern the spatial structure of the ion crystal. Control over the PES was achieved in our experiments by varying the ion trap parameters and by tuning the polarisability of ionic Rydberg states through microwave dressing. Our work opens new opportunities for the quantum simulation of molecular processes in regimes which are intractable by numerical simulations. In such a quantum simulator, the fundamental molecular shape is controlled by the number of ions and the geometry of the trapping field. The structure of the excited PES can be engineered by the interplay of external confinement, microwave control and dipolar interactions. Possible use cases of such a simulator are the investigation of non-adiabatic processes in the vicinity of crossing PES or the study of the impact of vibronic coupling on excitation transport across large molecules \cite{whitlow_quantum_2023,valahu_direct_2023,brox2017}. For the latter, one may exploit resonant dipole-dipole interactions among several ions excited to Rydberg states, as has been shown with neutral atoms \cite{Ravets2014Dec,deLeseleuc2019Aug}. A further avenue of investigation concerns the creation of superposition states of macroscopically different ion crystals \cite{feldker_mode_2014,Baltrusch2011Dec}. In the setup discussed in our work, this can in principle be achieved by preparing the central ion in a superposition between states of which one component is coupled to the Rydberg state and the other one is not. Subsequent excitation can then yield a macroscopic superposition state which may serve as a resource for addressing questions concerning quantum gravity and collapse models \cite{Carney2021Jan}. 

\section{Acknowledgement}
We gratefully acknowledge discussions with F. Schmidt-Kaler. This work was supported by the Swedish Research Council (Grant No. 2021-05811), and by the Knut \& Alice Wallenberg Foundation (Wallenberg Centre for Quantum Technology [WACQT]). W.L. acknowledges support from the EPSRC through Grant No. EP/W015641/1. This project has also received funding from the European Union’s Horizon Europe research and innovation programme under Grant Agreement No.\ 101046968 (BRISQ). This work is supported by the ERC grant OPEN-2QS (Grant No.\ 101164443, https://doi.org/10.3030/101164443).

\bibliography{bibliography}
\newpage
\section*{Supplemental Material}
\subsection*{Experimental setup}
\begin{figure}[t]
\centering
\includegraphics[scale=1]{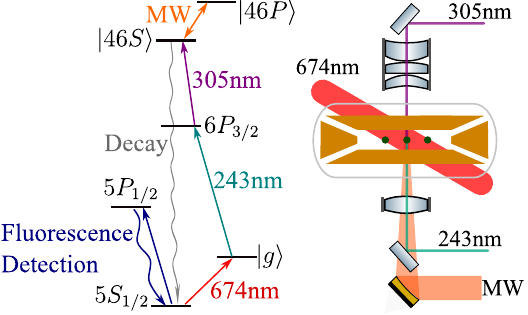}
\caption{\textbf{Level scheme and experimental setup.} Three ions are confined in a linear Paul trap (shown in orange). Coherent operations between the low-lying electronic states are driven globally by a laser at \SI{674}{\nano\meter}. Individual Rydberg excitation is performed by addressing the central ion using counter-propagating laser beams at \SI{305}{\nano\meter} and \SI{243}{\nano\meter}. A MW field, aligned parallel to the \SI{243}{\nano\meter} beam and perpendicular to the trap axis, is used to couple Rydberg states.}
\label{fig:setup}
\end{figure}

Spectroscopy scans of Rydberg resonances require a multitude of laser beams, aligned on the ion. A laser beam at \SI{674}{\nano\meter} was used to manipulate the low-lying electronic states $\ket{S}\equiv\ket{5\,^2S_{1/2}, m_J=-\frac{1}{2}}$ and $\ket{g}$. This beam is aligned at an angle of \SI{45}{\degree} to the longitudinal trapping axis. From the opposite direction, the cooling and fluorescence detection laser, and the repump lasers for the state preparation are aligned on the ion. To excite Rydberg states via an intermediate state, two ultraviolet (UV) beams at \SI{243}{\nano\meter} and \SI{305}{\nano\meter} wavelength are counter-propagating to minimise momentum transfer during the excitation. Both lasers are aligned in radial direction, addressing only the central ion of the chain. The microwave (MW) beam is also aligned on the ion from the radial direction. A level scheme and a sketch of the experimental setup can be found in Fig.~\ref{fig:setup}.

At the start of each experimental sequence, the ions are first cooled using Doppler cooling, followed by sideband cooling on all motional modes except the zigzag modes. Cooling of the latter one was not possible because the frequency was too close to the carrier transition; the sideband cooling process therefore led to off-resonant excitations. Next, all trap frequencies are characterised for the ground state. All ions are then excited from state $\ket{S}$ to state $\ket{g}$, followed by a post-selection step to ensure proper state transfer. Finally, the Rydberg resonance $\ket{g}\leftrightarrow\ket{r}$ is probed for the central ion, using a coherent spectroscopy technique, which relies on the Autler-Townes effect emerging in a three-level system, see \cite{higgins_quadrupol_2021,mokhberi_chapter_2020, bao_microwave-dressing_2025}.  Both, the intermediate state and the Rydberg state $\ket{r}$, decay with about 95\% probability to the lower-lying electronic state $\ket{S}$. This enables the following state-dependent fluorescence detection scheme to distinguish the long-lived states $\ket{g}$ and $\ket{S}$; when the ion is in state $\ket{g}$, the ion stays dark and no fluorescence is detected. If the population is coupled to the Rydberg state $\ket{r}$, it decays to the state $\ket{S}$. From there, the fluorescence transition $\ket{S}\leftrightarrow \ket{5{}^2P_{1/2}}$ is driven, the ion scatters photons, and a fluorescence signal is detected. Therefore, Rydberg excitation leads to photon counts, whereas no Rydberg excitation results in a dark signal. As the ion can decay with a probability of around $6\%$ to the dark $4D_{3/2}$ state, an additional repump laser is applied during the state-dependent fluorescence detection sequence.

\subsection*{Vanishing polarisability states}
The Rydberg states $\ket{46S}$ and $\ket{46P}$ have polarisabilities with opposite signs. Hence coupling them with a certain mixing ratio will make the total polarisability of the dressed state disappear \cite{pokorny2020}. The correct mixing ratio $\theta$ of the dressed states depends on the coupling strength $\Omega_\mathrm{MW}$ and the detuning of the coupling field $\Delta_\mathrm{MW}$. The total polarisability value is given by
\begin{align}\label{eq:total_pol}
    \mathcal{P}_\mathrm{r}=\mathcal{P}_\mathrm{46S}\cos^2(\theta)+\mathcal{P}_\mathrm{46P}\sin^2(\theta) \quad\text{with}\quad\tan(2\theta)=-\frac{\Omega_\mathrm{MW}}{\Delta_\mathrm{MW}}.
\end{align}
\begin{figure}[t]
\centering
\includegraphics[scale=1]{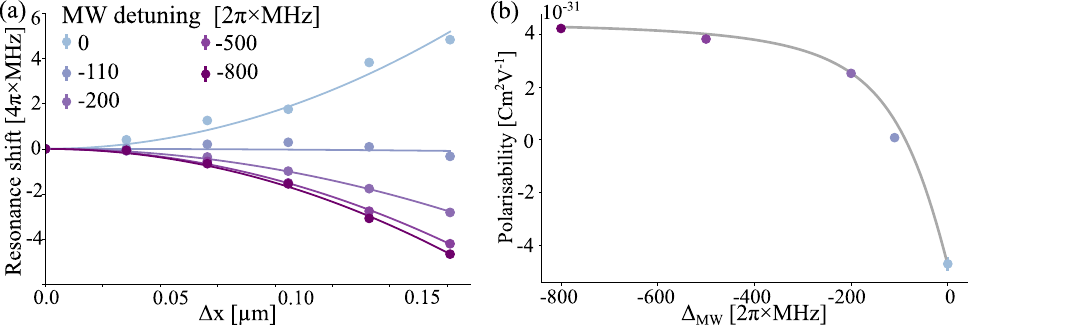}
\caption{\textbf{Polarisability-dependent Rydberg resonance shift.} For a fixed MW coupling strength between $\ket{46S}\leftrightarrow\ket{46P}$, the admixture of each state in the dressed state can be controlled via the MW detuning. (a) The polarisability is probed by measuring the resonance shift as a function of the ion's radial displacement from the trap centre. (b) The measured polarisability as a function of MW detuning is used to identify the detuning at which the total polarisability vanishes. For the dressed state used in the main text the detuning was set to -\SI{95}{\mega\hertz}, corresponding to a residual polarisability of around $0.048\mathcal{P}_\mathrm{46S}$.}
\label{fig:0pol}
\end{figure}
The polarisability value of the dressed state can be determined by measuring the effect of an offset electric field on the Rydberg excitation resonance. This offset electric field can easily be introduced by displacing the ion from the trap centre, which results in a mismatch between the RF and DC field nulls. The greater the total polarisability, the larger the resonance shift measured for a given displacement. The total polarisability value of the Rydberg state can be obtained by fitting the resonance shift using
\begin{align}\label{eq:pol_shift}
    \Delta\nu=-\frac{\mathcal{P}_\mathrm{r}\alpha^2\Delta x^2}{\hbar}.
\end{align}
However, as the dressed state approaches the vanishing polarisability condition, it becomes increasingly difficult to resolve the small resulting shifts in the Rydberg resonance. To determine the polarisability of the dressed state used in the measurements shown in the main text in Fig.~\ref{fig:Fig_2}(d), the following method was employed. 

Resonance shifts were measured as a function of increased radial displacements from the trap centre for various MW detunings, while keeping the coupling strength fixed at $\Omega_\mathrm{MW}=2\pi\times \SI{154}{MHz}$ (see Fig.~\ref{fig:0pol} (a)). The polarisability for each detuning was extracted by fitting the data using Eq.~(\ref{eq:pol_shift}). The resulting polarisability values as a function of the MW detuning are shown in Fig.~\ref{fig:0pol} (b). 

In the final analysis step, these values were fitted using Eq.~(\ref{eq:total_pol}) with two additional fit parameters. The first is a scaling parameter applied to the total polarisability, accounting for contributions from nearby Rydberg states not accounted in the two-level model. The second introduces a small shift to the MW detuning $\Delta_\mathrm{MW}$ to account for systematic experimental offsets. 

Based on the limited number of scans in Fig.~\ref{fig:0pol}, the MW detuning used for the measurements in the main text was fixed at $\Delta_\mathrm{MW}=-2\pi\times \SI{95}{\mega\hertz}$. According to the method described above, the residual polarisability at this detuning is approximately $0.048\mathcal{P}_\mathrm{46S}$. While this does not correspond to perfectly vanishing polarisability, it is sufficiently small to ensure $\omega_\mathrm{x,c}\approx\omega_\mathrm{x,c}^\mathrm{(r)}$ for the dressed state used for the measurements in the main text Fig.~\ref{fig:Fig_2}(d).

As discussed in the supplemental material of Ref.~\cite{zhang_submicrosecond_2020}, double ionisation becomes a significant problem when exciting an ion to Rydberg states in a room-temperature environment. When it occurs, the ion chain remain a doubly-charged ion, requiring the trap to be emptied and loaded with a new chain. Furthermore, all measurements presented in this work depend on the preparatory steps outlined above, which must be completed prior to determining the Rydberg resonance. As a result, the resonance can only be identified after all associated scans have been performed. 

\subsection*{Polarization dependent potential energy surfaces (PES)}
If the central ion of a three-ion crystal is excited to the Rydberg manifold, 
the pondermotive trap potential reads~\cite{jamesQuantumDynamicsCold1998,mullerTrappedRydbergIons2008,Weibin2011},
\begin{eqnarray}
&V&	= \frac{1}{2}  M\omega_\mathrm{x}^{(\text{r})2} x_1^2+\frac{1}{2} M\omega_\text{x}^2 \left(x_0^2+x_2^2\right)+\frac{1}{2}
	M\omega^2_\text{z}\left(z_0^2+z_1^2+z_2^2\right)+\\
	&&K_0\times\left[\frac{1}{\sqrt{(x_0-x_1)^2+(z_0-z_1)^2}}+\frac{1}{\sqrt{(x_0-{x_2})^2+({z_0}-{z_2})^2}}+\frac{1}{\sqrt{({x_2}-{x_1})^2+({z_2}-{z_1})^2}}\right], \nonumber
\end{eqnarray}
where $\omega_\text{x} = \sqrt{\frac{2e^2\alpha^2}{M^2\Omega_{RF}^2}-\frac{2e\beta}{M}}$ and $\omega_z = \sqrt{\frac{4e\beta}{M}}$ are the trap frequencies in the $x$ and $z$ direction. The parameters $\alpha$ and $\beta$ correspond to the electric field gradient in the radial and axial direction. $x_i$ and $z_i$ ($i=0,1,2$) are the coordinates of each ion, with $i=1$ denoting the central ion. The first term containing $\omega_\text{x}^{(\text{r})}$ describes the modified trapping potential in the Rydberg state and depends on the polarisability $\mathcal{P}_\mathrm{r}$. The trapping frequency in the Rydberg state can be approximated by $\omega_\text{x}^{(\text{r})}=\sqrt{\omega_\text{x}^2 - \frac{2e^2\alpha^2+4e^2\beta^2}{M}\mathcal{P}_\mathrm{r}}$. The parameter $K_0={e^2}/{4\pi\epsilon_0}$ with $e$ and $\epsilon_0$ to be the elementary charge and vacuum permittivity.

\subsubsection*{Equilibrium position and collective mode of the three-ion crystal}
The equilibrium positions and collective modes can be categorised into two regimes. One in which the ion chain forms a linear string, and one in which the ion crystal will arrange in a zigzag (ZZ) configuration. 

In the linear configuration, the ions are aligned along the $z$-axis. The equilibrium positions are $Z_2=-Z_0=(\frac{5K_0}{4\beta})^{1/3}$ and $Z_1=X_0=X_1=X_2=0$. The equilibrium position will not be affected by the Rydberg excitation. Using the equilibrium positions, we evaluate the Hessian matrix $M = \sum_{jk}\frac{\partial^2 V}{\partial r_j\partial r_k}$~\cite{jamesQuantumDynamicsCold1998}, 
\begin{equation}
	M=\left(\begin{array}{cccccc}
		\omega_\text{x}^2-\frac{9\mathcal{B}}{10m} & \frac{4\mathcal{B}}{5 m} & \frac{\mathcal{B}}{10 m} & 0 & 0 & 0 \\
		\frac{4\mathcal{B}}{5 m} & \mathcal{A}_1-\frac{8\mathcal{B}}{5m} & \frac{4 \mathcal{B}}{5 m} & 0 & 0 & 0 \\
		\frac{\mathcal{B}}{10 m} & \frac{4 \mathcal{B}}{5 m} & \omega_\text{x}^2-\frac{9\mathcal{B}}{10m} & 0 & 0 & 0 \\
		0 & 0 & 0 & \frac{14 \mathcal{B}}{5 m} & -\frac{8 \mathcal{B}}{5 m} & -\frac{\mathcal{B}}{5 m} \\
		0 & 0 & 0 & -\frac{8 \mathcal{B}}{5 m} & \frac{21 \mathcal{B}}{5 m} & -\frac{8 \mathcal{B}}{5 m} \\
		0 & 0 & 0 & -\frac{\mathcal{B}}{5 m} & -\frac{8 \mathcal{B}}{5 m} & \frac{14 \mathcal{B}}{5 m} \\
	\end{array}\right),
\end{equation}
where $\mathcal{A}_1=\omega_\text{x}^2+A_\text{e}/m \approx \omega_\text{x}^2 +  2e^2\alpha^2\mathcal{P}_\mathrm{r}/m$. We have defined $A_\text{e}=2e^2\alpha^2\mathcal{P}_\mathrm{r}$ and $\mathcal{B}=4e\beta$ and used the fact that $\alpha\gg \beta$ in the Paul trap. The Hessian matrix $M$ is block diagonal, where the upper (lower) block corresponds to vibrations along the $x$-axis ($z$-axis). By diagonalizing the block matrix, we obtain three eigenmodes in each axis. In the linear crystal, the axial modes (modes along the $z$-axis) are not affected by the Rydberg excitation. In the following analysis, we will focus on the radial modes. The corresponding Hessian matrix $M_{\text{L}}$ is given by the upper block of $M$.

We first find the phonon mode without carrying out the Rydberg excitation. By diagonalising the block matrix $M_{\text{L}}$, the eigenvectors can be obtained, which form matrix $U_{\text{L}}$,
\begin{equation}
    U_\text{L}=\left(\begin{array}{ccc}
    \frac{1}{\sqrt{3}} & \frac{1}{\sqrt{3}} & \frac{1}{\sqrt{3}} \\
    \frac{1}{\sqrt{6}} & -\sqrt{\frac{2}{3}} & \frac{1}{\sqrt{6}} \\
    -\frac{1}{\sqrt{2}} & 0 & \frac{1}{\sqrt{2}}\end{array}\right),
\end{equation}
where the first row is the center-of-mass (CM) mode, the second row is the ZZ, and the last row the rocking modes. The mode frequency can be obtained via the canonical transformation, 
\begin{eqnarray}
    U_\text{L}M_\text{L}U_\text{L}^{T} = \left( \begin{array}{ccc} \omega_\text{x}^2 &0& 0 \\ 
    0 & \omega_\text{x}^2 -\frac{12\mathcal{B}}{5m} &0 \\
     0& 0 & \omega_\text{x}^2 -\frac{\mathcal{B}}{M}\end{array} \right).
\end{eqnarray}
The square root of the diagonal matrix element gives the corresponding mode frequency. 

When the central ion is excited to the Rydberg state, the Hessian matrix $M_\text{L}^{(\text{r})}$ can not be fully diagonalised using $U_{\text{L}}$. After carrying out the canonical transformation, we obtain,
\begin{equation}
     U_\text{L}M_\text{L}^{(\text{r})} U_\text{L}^{T}=\left(\begin{array}{ccc}\omega_{\text{x}}^2+\frac{A_\text{e}}{3m} & -\frac{\sqrt{2}}{3}A_\text{e} & 0 \\
    -\frac{\sqrt{2}}{3}A_\text{e} & \omega_\text{x}^2 +\frac{10A_\text{e}-36\mathcal{B}}{15m} & 0 \\
    0 & 0 & \omega_\text{x}^2 -\frac{\mathcal{B}}{m}\end{array}\right).
    \label{eq:MLr}
\end{equation}
This shows that the rocking mode is not affected by the Rydberg excitation. We will exclude it in the following discussion. 

Looking at the upper $2\times 2$ block in Eq.~(\ref{eq:MLr}), the diagonal matrix elements are modified by the Rydberg excitation due to $A_{\text{e}}\neq 0$. In the Rydberg $nS$ states, the polarisability $\mathcal{P}_\mathrm{r}<0$. Hence the mode frequencies are lowered by the Rydberg excitation. The CM and ZZ modes (as defined by $U_{\text{L}}$) are coupled with coupling strength $-\sqrt{2}A_{\text{e}}/3$. If we diagonalise $M_{\text{L}}^{(\text{r})}$, the resulting two phonon modes are different from the mode vector in $U_{\text{L}}$. We call them modified CM and ZZ modes. 

In the zigzag configuration, the ions will be pushed away from $X_j=0$. The equilibrium positions are evaluated analytically, 
\begin{eqnarray*}
Z_0&=&-\left[\frac{K_0}{4\beta -\frac{4\omega_\text{x}^2\mathcal{A}_1}{2\omega_\text{x}^2 + \mathcal{A}_1}}\right]^{1/3}, \\
Z_1 &=& 0, \\
Z_2 &=& -Z_0,
\end{eqnarray*}
in the axial direction and 
\begin{eqnarray*}
X_0 &=& -\frac{\mathcal{A}_1}{2\omega_\text{x}^2 + \mathcal{A}_1}\sqrt{\left[\frac{K_0(2\omega_\text{x}^2+\mathcal{A}_1)}{\omega_\text{x}^2\mathcal{A}_1}\right]^{2/3}-Z_0^2},\\
X_1 &=& -\frac{2\omega_\text{x}^2}{\mathcal{A}_1}X_0, \\
X_2 &=& X_0,
\end{eqnarray*}
in the radial direction.

Note that the equilibrium position now explicitly depends on the extra potential ($A_{\text{e}}/m$) induced by the Rydberg excitation. This is different from the linear case, where only the trap frequency is modified by the Rydberg excitation. Moreover, all phonon modes but the CM mode along the $z$-axis consist of vibrations along both the $x$- and $z$-axis. This is an important difference compared to the linear regime. Properties of these modes (eigenvectors and eigenvalues) are calculated numerically.

\subsection*{Rydberg excitation in the linear configuration}
In case the Rydberg excitation does not alter the equilibrium positions of the ion crystal, the collective modes along the axial axis will not be affected. Without the Rydberg excitation, the two relevant modes are the CM and ZZ mode, whose Hamiltonian is given by
\begin{eqnarray}
    H_\text{g} = \omega_{\text{cm}} a_1^{\dagger}a_1 +\omega_{\text{zz}}a_2^{\dagger}a_2 + \frac{\omega_{\text{cm}}+\omega_{\text{zz}}}{2},
\end{eqnarray}
with $\omega_{\text{cm}} = \omega_\text{x}$ and $\omega_{\text{zz}} = \sqrt{\frac{2e^2\alpha^2}{M^2\omega_\text{x}^2}-\frac{58}{5}\frac{e\beta}{M}}$. We have defined ${a}_j ({a}_j^{\dagger}$) as the annihilation (creation) operators of the two modes.

If the central ion is excited to the Rydberg state, the Hamiltonian of the modified CM and ZZ modes is given by
\begin{equation}
    H_{\text{r}}= {\omega}_1^{(\text{r})}{b}^{\dagger}_1{b}_1 + {\omega}_2^{(\text{r})}{b}_2^{\dagger}{b}_2,
\end{equation}
where $\omega_j^{(\text{r})}$ ($j=1,2$) are the mode frequencies, and $\tilde{b}_j (\tilde{b}_j^{\dagger}$) are the annihilation (creation) operators. Since the modes of the LLE and excited state are not orthogonal, the laser coupling of the electronic states depends on the overlap $C_{\mathbf{n}}^{\mathbf{m}}$ of phonon wave functions on the lower and upper PES. Here $\mathbf{n}=\{n_{\text{cm}},n_{\text{zz}}\}$ and $\mathbf{m}=\{m_1,m_2\}$ denoting the Fock states of the two modes of the corresponding PES. The FC factors can be obtained from the overlap, i.e. $|C_{\mathbf{n}}^{\mathbf{m}}|^2$. 
\begin{figure}[t]
\centering
\includegraphics[width=0.95\textwidth]{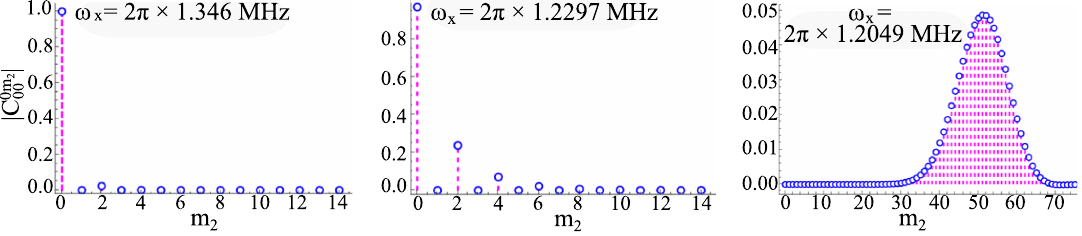}
\caption{\textbf{Franck-Condon factors.} Phonon states in the lower PES are in the vacuum state ($n_{\text{cm}}=n_{\text{zz}}=0$). We vary phonon number $m_2$ of the modified ZZ mode. Away from the critical point ($\omega_\text{x} = 2\pi\times 1.346$ MHz), the FC factors have a narrow distribution. Close to the critical point ($\omega_\text{x} = 2\pi\times1.234$ MHz), the ion crystal before and after the Rydberg excitation is still in a linear configuration. However the mode becomes soft, such that the even phonon number states couple to the vacuum state of the CM and ZZ mode on the lower PES. When $\omega_\text{x} = 2\pi\times 1.225$ MHz, the Rydberg excitation changes the crystal from the linear to ZZ configuration. The displacement of the equilibrium position along the $x$-axis means the vacuum state on the ground state PES can only couple to very high phonon number states on the Rydberg state PES. In this example, we find the FC factors has a broad distribution centred around $m_2\approx 50$. }
\label{fig:OmegaGS0}
\end{figure}

The overlap $C_{\mathbf{n}}^{\mathbf{m}}$ are calculated using the Duschinsky transformation~\cite{BorrelliRaffaele2013Feb}. They are high dimensional coefficients, depending on the phonon modes on the two PESs. Some examples are plotted in Fig.~\ref{fig:OmegaGS0}. For the numerical simulation we assume that the two phonon modes on the lower PES are in the vacuum state ($n_{\text{cm}}=0,n_{\text{zz}}=0$). Then we vary phonon state ($m_2$) of the modified ZZ mode (with $m_1=0$). When the trapping potential deformation is negligible, the modes on the two PESs are largely identical. The FC factor has a single peak at $m_2=0$ (see Fig.~\ref{fig:OmegaGS0}, $\omega_{\text{x}}=2\pi\times 1.346$ MHz). When the trapping potential along the $x$-axis is weakened by the Rydberg excitation, the phonon modes on the upper PES become soft, and are not orthogonal to the modes on the lower PES. This allows the vacuum state to couple to different phonon number states (see the case $\omega_{\text{x}}=2\pi\times 1.234$ MHz, Fig.~\ref{fig:OmegaGS0}). Note that only even number Fock states are coupled due to the parity symmetry of the FC coupling. On the other hand, the phonon modes on the two PESs are so different when a conformational change takes place. Not only the mode frequencies differ significantly, but also the centres of the phonon wave packets are displaced. Here we show the example with $\omega_\text{x}=2\pi\times 1.225$ MHz. Without the Rydberg excitation, the crystal is in a linear configuration. After the central ion is excited to the Rydberg state, it changes to a ZZ configuration. Due to the displacement of the potential minima and modification of the trap frequencies, the FC factor has a broad distribution. Rydberg excitation is reduced significantly due to the vanishing phonon coupling, even when the laser is resonant with the electronic transition.

\subsection*{Numerical simulation of the Rydberg excitation} 
In the experiment, phonons are in thermal states initially. The central ion is laser excited from the metastable state $\ket{g}=\ket{4D_{5/2}}$ to Rydberg state $\ket{r}=\ket{46S}$ via intermediate state $\ket{p}=\ket{6P_{3/2}}$. The Rydberg excitation is realised through a two-photon transition with Rabi frequencies $\Omega_\mathrm{gp}$ and $\Omega_\mathrm{pr}$, respectively. The Rydberg ion decays to state $\ket{s}=\ket{5S_{1/2}}$ incoherently, and subsequently is detected through fluorescence. Details of the experiment are described in Sec. I. Motivated by the experiment, we model the two-photon coupling with Hamiltonian,
\begin{equation}
	H_\text{s} = \delta \sigma_\mathrm{rr} + \frac{\Omega_\mathrm{gp}}{2}(\sigma_\mathrm{gp}+\sigma_\mathrm{pg}) +\frac{\Omega_\mathrm{pr}}{2}(\sigma_\mathrm{pr}+\sigma_\mathrm{rp}),
\end{equation} 
where $\delta$ and $\Omega_j$ ($j=\text{gp}$, and $\text{pr}$) are the detuning and Rabi frequencies of the lower and upper coupling. Transition operators, $\sigma_{jk} = |j\rangle\langle k|$, are defined using electronic states $|j\rangle$ ($j=$s, g, p, and $\text{r}$). In the electronically low-lying states $|s\rangle$, $|g\rangle$ and $|p\rangle$, the phonon modes are described by Hamiltonian $H_\text{g}$. In the Rydberg state $|r\rangle$, the phonon Hamiltonian is $H_\text{r}$. The total Hamiltonian of the system is $H_\text{o}=\mathbf{I}_\text{p}\otimes H_\text{s} + H_\text{g}\otimes (\textbf{I}_\text{e}-\sigma_{\text{rr}}) +H_\text{r}\otimes \sigma_{\text{rr}}$. Here $\textbf{I}_\text{p}$ is the phonon identity operator, and $\textbf{I}_\text{e} = \sum_{j}  \sigma_{jj}$ is the identity operator of the electronic state.

\begin{figure}[t]
	\centering
\includegraphics[width=0.55\textwidth]{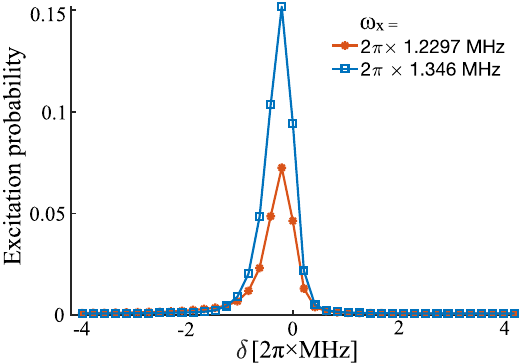}
	\caption{\textbf{Rydberg population as a function of $\delta$.} Both cases correspond to a situation in which the Rydberg excitation does not induce a conformational change. Close to the critical point, the Rydberg population decreases as the mode becomes soft. In the simulation, the decay rate to state $|g\rangle$ is $\gamma_\text{gp} = 25$ MHz, and Rydberg lifetime $\tau = 5.5\,\mu$s. The lower and upper transition Rabi frequencies are $\Omega_\mathrm{gp} = 2\pi\times 1.0 $ MHz and $\Omega_\mathrm{pr}=2\pi\times 1.5 $ MHz.}
	\label{fig:compare_dynamics}
\end{figure}
Taking into account the spontaneous decay in the excited states, we study the laser-ion coupling dynamics with the Lindblad 
master equation $\dot{\rho}=-i[H_\text{o},\rho] +\sum_j\mathcal{D}_j(\rho)$, where the decay of the electronic states is described by the operators $\mathcal{D}_j(\rho)=  \gamma_j \left(s_j\rho s_j^{\dagger}+\frac{1}{2}\{s_j^{\dagger}s_j,\rho\}\right)$ with jump operators $s_j$ and rates $\gamma_j$. Here we include three decay processes, $|p\rangle \to |s\rangle$, $|p\rangle \to |g\rangle$ and $|r\rangle \to |s\rangle$, described by jump operators $s_\mathrm{sp} = |s\rangle \langle p|$, $s_\mathrm{gp} = |g\rangle \langle p|$ and $s_\mathrm{sr}=|s\rangle\langle r|$, respectively. In the numerical simulation, the ion is intitialised in state $|g\rangle$ at $t=0$, and the phonon number state has a thermal distribution, $\rho_p=\frac{1}{n_p+1}\sum_{m=0}^{\infty} \left(\frac{n_p}{n_p+1}\right)^m$ with $n_p$ the mean phonon number of the $p$-th mode.

In Fig.~\ref{fig:compare_dynamics}, we show Rydberg population for different trap frequencies $\omega_\text{x}$. It can be found that there is a peak in the Rydberg population around the resonance $\delta\approx 0$. The peak is slightly red-shifted, due to the trapping potential becoming shallower in the Rydberg state. 
When comparing the two cases, the maximal Rydberg population decreases when $\omega_\text{x}$ approaches the critical value. 
This is a direct consequence of the Franck-Condon effect. Close to the conformational transition, the PES in the ground state and in the Rydberg state are so different, such that the phonon coupling, determined by the Franck-Condon coefficients, becomes smaller. This strongly reduces the Rydberg excitation, which has been observed in our experiment. 

\end{document}